# Quantum Circuits: Fanout, Parity, and Counting

Cristopher Moore[1]

Santa Fe Institute, 1399 Hyde Park Road, Santa Fe, New Mexico 87501
moore@santafe.edu

**Abstract.** We propose definitions of $\mathbf{QAC}^0$, the quantum analog of the classical class $\mathbf{AC}^0$ of constant-depth circuits with AND and OR gates of arbitrary fan-in, and $\mathbf{QACC}^0[q]$, where $n$-ary $\mathrm{MOD}_q$ gates are also allowed. We show that it is possible to make a 'cat' state on $n$ qubits in constant depth if and only if we can construct a parity or $\mathrm{MOD}_2$ gate in constant depth; therefore, any circuit class that can fan out a qubit to $n$ copies in constant depth also includes $\mathbf{QACC}^0[2]$. In addition, we prove the somewhat surprising result that parity or fanout allows us to construct $\mathrm{MOD}_q$ gates in constant depth for any $q$, so $\mathbf{QACC}^0[2] = \mathbf{QACC}^0$. Since $\mathbf{ACC}^0[p] \ne \mathbf{ACC}^0[q]$ whenever $p$ and $q$ are mutually prime, $\mathbf{QACC}^0[2]$ is strictly more powerful than its classical counterpart, as is $\mathbf{QAC}^0$ when fanout is allowed.

## 1 Introduction

The theory of circuit complexity has become an important branch of theoretical computer science. Shallow circuits correspond to parallel algorithms that can be performed in small amounts of time on an idealized parallel computer, and so circuit complexity can be thought of as a study of how to solve problems in parallel. In addition, some low-lying circuit classes have beautiful algebraic characterizations, e.g. [3,4,9].

In [11,12], Moore and Nilsson suggested a definition of $\mathbf{QNC}$, the quantum analog of $\mathbf{NC}$, the class of problems solvable by circuits with polylogarithmic depth and polynomial size. Here, we will study quantum versions of some additional circuit classes. Recall the following definitions:

1. $\mathbf{NC}^k$ consists of problems solvable by families of circuits with depth $\mathcal{O}(\log^k n)$ and size polynomial in $n$, where $n$ is the size of the input, where the allowed gates are ANDs and ORs with two inputs (which may be negated).
2. $\mathbf{AC}^k$ is like $\mathbf{NC}^k$, but where we allow AND and OR gates with unbounded fan-in, i.e. arbitrary numbers of inputs, in each layer of the circuit.
3. $\mathbf{ACC}^k[q]$ is like $\mathbf{AC}^k$, but where we also allow $\mathrm{MOD}_q$ gates with unbounded fan-in, where $\mathrm{MOD}_q(x_1, \ldots, x_n)$ outputs 1 iff the sum of the inputs is not a multiple of $q$.
4. $\mathbf{ACC}^k = \cup_q \mathbf{ACC}^k[q]$.
5. $\mathbf{NC} = \cup_k \mathbf{NC}^k = \cup_k \mathbf{AC}^k = \cup_k \mathbf{ACC}^k$.

Then we have

$$\mathbf{AC}^0 \subset \mathbf{ACC}^0[2] \subset \mathbf{ACC}^0 \subseteq \mathbf{NC}^1 \subseteq \cdots \subseteq \mathbf{NC}$$

In fact, these first two inclusions are known to be proper [1,7,14,19]. Neither MAJORITY nor PARITY are in $\mathbf{AC}^0$, while the latter is trivially in $\mathbf{ACC}^0[2]$. In addition, $\mathbf{ACC}^0[p]$ and $\mathbf{ACC}^0[q]$ are known to be incomparable whenever $p$ and $q$ are mutually prime. Thus these classes give us some of the few strict inclusions known in computational complexity theory.

There is significant evidence that quantum complexity classes are often more powerful than their classical counterparts. The best-known is Shor's factoring algorithm [17], showing that integer factoring is in $\mathbf{BQP}$, the quantum analog of bounded-probability polynomial time. Other results lower down on the computational hierarchy include the recognition of non-regular languages by two-way quantum finite automata with bounded probability [8] and quantum context-free languages that are not classically context-free [10].

In this paper, we prove a number of results about $\mathbf{QAC}$ and $\mathbf{QACC}$, and address some definitional difficulties. We show that an ability to form a 'cat state' with $n$ qubits, or fan out a qubit into $n$ copies in constant depth, is equivalent to being able to constrict an $n$-ary parity gate in constant depth. Furthermore, we show the somewhat surprising result that either of these allows us to build $\text{MOD}_q$ gates in constant depth for any $q$, so $\mathbf{QACC}^0[2] = \mathbf{QACC}^0$. This is markedly different from the classical situation mentioned above. Finally, we discuss how best to compare these circuit classes to classical ones, and conclude that $\mathbf{QACC}^0[2]$, and a form of $\mathbf{QAC}^0$ we call $\mathbf{QAC}^0_{\text{wf}}$, are strictly more powerful than $\mathbf{ACC}^0[2]$ and $\mathbf{AC}^0$.

## 2 Definitions

We will use the notation in figure 1 for our various gates. The $n$-ary Toffoli gate, or simply the AND gate, negates the target qubit if all the input qubits are true. The $n$-ary controlled-$U$ gate applies a unitary transformation $U$ to the target if all the inputs are true, and the $\text{MOD}_q$ gate negates the target if the number of true inputs is not a multiple of $q$. Note that all these are reversible. For three inputs, for instance, these matrices can be written

$$\begin{pmatrix} 1 & & & & & & & \\ & 1 & & & & & & \\ & & 1 & & & & & \\ & & & 1 & & & & \\ & & & & 1 & & & \\ & & & & & 1 & & \\ & & & & & & 1 & \\ & & & & & & & X \end{pmatrix}, \quad \begin{pmatrix} 1 & & & & & & & \\ & 1 & & & & & & \\ & & 1 & & & & & \\ & & & 1 & & & & \\ & & & & 1 & & & \\ & & & & & 1 & & \\ & & & & & & 1 & \\ & & & & & & & U \end{pmatrix}, \quad \text{and} \quad \begin{pmatrix} 1 & & & & & & & \\ & X & & & & & & \\ & & X & & & & & \\ & & & 1 & & & & \\ & & & & X & & & \\ & & & & & 1 & & \\ & & & & & & 1 & \\ & & & & & & & X \end{pmatrix}$$

(this last is a $\text{MOD}_2$ gate) where $\mathbf{1} = \begin{pmatrix} 1 & 0 \\ 0 & 1 \end{pmatrix}$ and $X = \begin{pmatrix} 0 & 1 \\ 1 & 0 \end{pmatrix}$. We will also use the symmetric phase shift gate, shown here for two qubits $\begin{pmatrix} 1 & 0 & 0 & 0 \\ 0 & 1 & 0 & 0 \\ 0 & 0 & 1 & 0 \\ 0 & 0 & 0 & e^{i\theta} \end{pmatrix}$ which is a special case of the controlled-$U$ gate, and the one-qubit *Hadamard gate* $H = \frac{1}{\sqrt{2}} \begin{pmatrix} 1 & 1 \\ 1 & -1 \end{pmatrix}$. Note that the controlled $\pi$-shift is simply the controlled-not with the target qubit conjugated with $H$, and that $H^2 = \mathbf{1}$. Finally, we will also sometimes use controlled-$U$ gates with some inputs negated, which can be performed by conjugating that input with $X$. All of these are shown in figure 1.

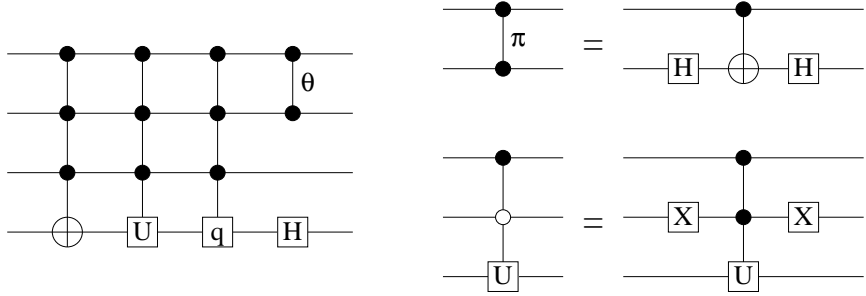

**Fig. 1.** Our notation for $n$-ary Toffoli, controlled-$U$, and $\text{MOD}_q$ gates, fanout gates, symmetric phase shift gates, and the Hadamard gate. On the top right, we show a useful identity between the controlled-not, the controlled $\pi$-shift, and the Hadamard gate. On the bottom right we show a controlled-$U$ gate with one of its inputs negated by conjugation with $X$.

Then our definition of **QAC** and **QACC** is as follows:

**Definition.** A family of unitary operators $F_n \in U(2^n)$ is in $\mathbf{QAC}^k$ if $F_n$ can be written as a product of $\mathcal{O}(\log^k n)$ operators, each of which is a tensor product of one-qubit gates and Toffoli gates acting on disjoint sets of qubits. Each such tensor product is called a *layer*. The definition of $\mathbf{QACC}^k[q]$ is similar, except that $\text{MOD}_q$ gates are also allowed. Finally, $\mathbf{QACC}^k = \cup_q \mathbf{QACC}^k[q]$.

We also allow our circuits to use a polynomial number of *ancillae* or work qubits. However, for the final measurement to be correct, these ancillae have to be unentangled from the qubits of the final state that we care about. Therefore, we require that the ancillae start and end in a pure state $|0\rangle$, in which case our desired operator occurs as a diagonal block of the larger operator on this subspace [11]. This also allows us to re-use ancillae.

Note that $n$-ary controlled-$U$ gates can be performed in constant depth with one ancilla, by writing them as a product of two Toffoli gates and a two-qubit controlled-$U$ gate with one input, which can then be written as a product of controlled-nots and one-qubit gates [2]. Thus our definition of **QAC** remains the same if we allow $n$-ary controlled-$U$ gates in each layer. Similarly, allowing $\text{MOD}_q$ controlled-$U$ gates, which apply $U$ to the target qubit if the number of true inputs is not a multiple of $q$, does not change our definition of **QACC**.

It is, of course, tempting to pronounce these classes 'quack.'

## 3  Fanout, Cat states and parity

To make a shallow parallel circuit, it is often important to *fan out* one of the inputs into multiple copies. One of the differences between classical circuits and quantum ones as we have defined them here is that in classical circuits, we usually assume that we get arbitrary fanout for free, simply by splitting a wire into as many copies as we like. This is difficult in quantum circuits, since making an unentangled copy requires non-unitary, and in fact non-linear, processes since

$$(\alpha|0\rangle + \beta|1\rangle) \otimes (\alpha|0\rangle + \beta|1\rangle) = \alpha^2|00\rangle + \alpha\beta(|01\rangle + |10\rangle) + \beta^2|11\rangle$$

has coefficients quadratic in $\alpha$ and $\beta$. This is the so-called 'no cloning' theorem.

However, the controlled-not gate can be used to copy a qubit onto an ancilla in the pure state $|0\rangle$ by making a non-destructive measurement:

$$(\alpha|0\rangle + \beta|1\rangle) \otimes |0\rangle \rightarrow \alpha|00\rangle + \beta|11\rangle$$

Note that the final state is not a tensor product of two independent qubits, since the two qubits are completely entangled. This means that whatever we do to one copy, we do to the other. Except when the states are purely Boolean, we have to treat this kind of 'fanout' more gingerly than we would in the classical case.

By making $n$ copies of a qubit in this sense, we can make a 'cat state' $\alpha|000\cdots0\rangle + \beta|111\cdots1\rangle$. Such states are useful in making quantum computation fault-tolerant (e.g. [6,18]). We can do this in $\log n$ depth with controlled-not gates, as shown in the left-hand part of figure 2. When preceded by a Hadamard gate on the top qubit, this will map an initial state $|0000\rangle$ onto a cat state $\frac{1}{\sqrt{2}}(|0000\rangle + |1111\rangle)$. However, we will also consider circuits which can do this in a single layer, with a 'fanout gate' that simultaneously copies a qubit onto $n$ target qubits. This is simply the product of $n$ controlled-not gates, as shown in the right-hand part of figure 2.

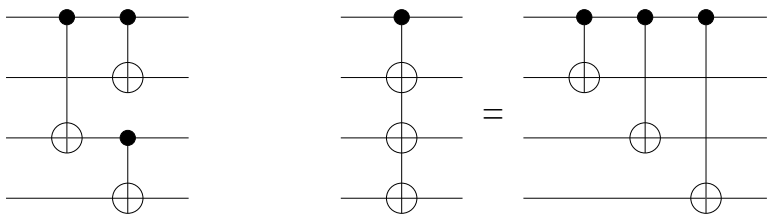

**Fig. 2.** Two ways to make a cat state on $n$ qubits. The circuit on the left uses only two-qubit gates and has depth $\log n$. On the right, we define a 'fanout gate' that simultaneously performs $n$ controlled-nots from one input qubit.

We now show that in quantum circuits, we can do fanout in constant depth if and only if we can construct a parity gate in constant depth.

**Proposition 1.** *In any class of quantum circuits that includes Hadamard and controlled-not gates, the following are equivalent:*

1. *It is possible to map $\alpha|0\rangle + \beta|1\rangle$ and $n-1$ ancillae in the state $|0\rangle$ onto an $n$-qubit cat state $\alpha|000\cdots 0\rangle + \beta|111\cdots 1\rangle$ in constant depth.*
2. *The n-ary fanout gate of figure 2 can be implemented in constant depth with at most $n-1$ additional ancillae.*
3. *An n-ary parity or $MOD_2$ gate as defined above can be implemented in constant depth with at most $n-1$ additional ancillae.*

*Proof.* First, note that (1) is *a priori* weaker than (2), since (1) only requires that an operator map $|000\cdots 0\rangle$ to $|000\cdots 0\rangle$ and $|100\cdots 0\rangle$ to $|111\cdots 1\rangle$. In fact, the two circuits shown in figure 2 both do this, even though they differ on other initial states.

To prove $(2 \Leftrightarrow 3)$, we simply need to notice that the parity gate is a fanout gate going the other way conjugated by a layer of Hadamard gates, since parity is simply a product of controlled-nots with the same target qubit, and conjugating with $H$ reverses the direction of a controlled-not. This is shown in figure 3. Clearly the number of ancillae used to perform either gate will be the same.

To prove $(1 \Rightarrow 3)$, we use a slightly more elaborate circuit shown in figure 4. Here we use the identity shown in figure 1 to convert the parity gate into a product of controlled $\pi$-shifts. Since these are diagonal, they can be parallelized as in [11] by copying the target qubit onto $n-1$ ancillae, and applying each one to a different copy. While we have drawn the circuit with a fanout gate, any gate that satisfies the conditions in (1) will do.

Finally, $(2 \Rightarrow 1)$ is obvious. □

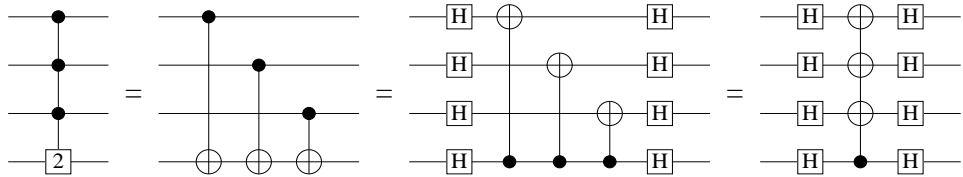

**Fig. 3.** The parity and fanout gates are conjugates of each other by a layer of Hadamard gates.

This brings up an interesting issue. It is not clear that $\mathbf{AC}^0 \subset \mathbf{QAC}^0$ as we have defined $\mathbf{QAC}^0$ here, since we are not allowing arbitrary fanout in each layer. An alternate definition, which we might call $\mathbf{QAC}$ *with fanout* or $\mathbf{QAC}_{\mathrm{wf}}$, would allow us to perform controlled-$U$ gates or Toffoli gates in the same layer whenever they have different target qubits, even if their input qubits overlap. This seems reasonable, since these gates commute. Since we can fanout to $n$ copies in $\log n$ layers as in figure 2, we have $\mathbf{QAC}^k \subseteq \mathbf{QAC}_{\mathrm{wf}}^k \subseteq \mathbf{QAC}^{k+1}$. We can define $\mathbf{QACC}_{\mathrm{wf}}$ in the same way, and Proposition 1 implies that $\mathbf{QAC}^k_{\mathrm{wf}} = \mathbf{QACC}^k_{\mathrm{wf}}[2] = \mathbf{QACC}^k[2]$.

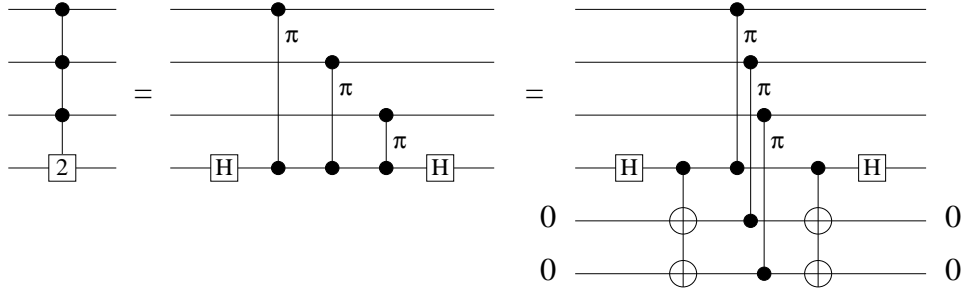

**Fig. 4.** The parity gate can also be written as a product of controlled $\pi$-shifts, with the target qubit conjugated by $H$. Since these are diagonal, we can parallelize them using any gate that can make a cat state.

We leave it to the reader to decide which definition, $\mathbf{QAC}^0$ or $\mathbf{QAC}^0_{\mathrm{wf}}$, is a better analog of $\mathbf{AC}^0$.

## 4 Parity or fanout gives $\mathrm{MOD}_q$ gates

As stated above, in the classical case $\mathrm{MOD}_p$ and $\mathrm{MOD}_q$ gates are not easy to build from each other whenever $p$ and $q$ are relatively prime. In fact, to do it in constant depth requires a circuit of exponential size [19]. In this section, we will show this is not true in the quantum case. Specifically, with $n$-ary parity gates we can build $\mathrm{MOD}_q$ gates in constant depth.

**Proposition 2.** *In any circuit class containing $n$-ary parity gates and one-qubit gates, we can construct an $n$-ary $\mathrm{MOD}_q$ gate, with $n \lceil \log q \rceil$ ancillae, in depth depending only on $q$.*

*Proof.* Let $k = \lceil \log q \rceil$, and let $M$ be a Boolean matrix on $k$ qubits where the zero state has period $q$. For instance, if we write $|x\rangle$ for $0 \leq x < 2^k$ as shorthand for $|x_{k-1} \cdots x_1 x_0\rangle$ where the $x_i$ are the digits of $x$'s binary expansion, then let $M$ permute the $|x\rangle$ as follows:

$$M|x\rangle = \begin{cases} |(x+1) \bmod q\rangle & \text{if } x < q \\ |x\rangle & \text{if } x \geq q \end{cases}$$

Then if we start with $k$ ancillae in the state $|0\rangle$ and apply a controlled-$M$ gate to them from each input, the state will differ from $|0\rangle$ on at least one qubit if and only if the number of true inputs is not a multiple of $q$. (Note that this controlled-$M$ gate applies to $k$ target qubits at once in an entangled way.) We can then apply an $n$-ary OR to these $k$ qubits, i.e. a Toffoli gate with its inputs conjugated with $X$ and its target qubit negated before or after the gate, and return the $k$ ancillae to $|0\rangle$ by applying the inverse series of controlled-$M^\dagger$ gates.

Now we use Proposition 4 of [11] to parallelize this set of controlled-$M$ gates. We can convert them to diagonal gates by conjugating the target qubits with a unitary operator $T$, where $T^\dagger D T = M$ and $D$ is diagonal. If we have a parity gate, we can fan out the $k$ ancillae to $n$ copies each using Proposition 1. We can then simultaneously apply the $n$ controlled-$D$ gates from each input to the corresponding copy, and then uncopy them back.

This is shown in figure 5. For $q = 3$, for instance, we have $M = \begin{pmatrix} 0 & 1 & 0 & \\ 0 & 0 & 1 & \\ 1 & 0 & 0 & \\ & & & 1 \end{pmatrix}$, $T = \frac{1}{\sqrt{3}} \begin{pmatrix} 1 & 1 & 1 & \\ e^{4\pi i/3} & e^{2\pi i/3} & 1 & \\ e^{2\pi i/3} & e^{4\pi i/3} & 1 & \\ & & & \sqrt{3} \end{pmatrix}$, and $D = \begin{pmatrix} 1 & & & \\ & 1 & & \\ & & e^{2\pi i/3} & \\ & & & e^{4\pi i/3} \end{pmatrix}$.

The operators $T$, $T^\dagger$, and the controlled-$D$ gate can be carried out in some finite depth by controlled-nots and one-qubit gates by the results of [2]. The total depth of our $\text{MOD}_q$ gate is a function of these and so of $q$, but not of $n$. Finally, the number of ancillae used is $kn = n \lceil \log q \rceil$ as promised. $\square$

To look more closely at the depth as a function of $q$, we note that using the methods of Reck et al. [15] and Barenco et al. [2], any operator on $k$ qubits can be performed with $\mathcal{O}(k^3 4^k)$ two-qubit gates. Since $k = \lceil \log q \rceil$, this means that the depths of $T$, $T'$ and the controlled-$D$ are at most $\mathcal{O}(q^2 \log^3 q)$.

Since we can construct $\text{MOD}_q$ gates in constant depth, we have $\mathbf{QACC}^k[q] \subset \mathbf{QACC}^k[2]$ for all $q$, so $\mathbf{QACC}^k = \mathbf{QACC}^k[2]$. By proposition 1, these are both also equal to $\mathbf{QAC}^k_{\text{wf}}$. In particular, we have

$$\mathbf{QAC}^0_{\text{wf}} = \mathbf{QACC}^0[2] = \mathbf{QACC}^0$$

while classically all of these are strict inclusions.

To compare these with classical circuits requires a little care. For any Boolean circuit with $n$ inputs, $m$ outputs, depth $d$, and width $w$, it is easy to construct a reversible circuit on $n + m + k$ bits of depth $2d - 1$ where $n$ bits keep the input, $m$ bits get xor'ed with the output, and $k = wd$ ancillae start and end in the zero state. We do this by assigning an ancilla to each gate in the original circuit, and replacing each gate with a reversible one that xors that ancilla with the output. Then we can erase the ancillae by moving backward through the layers of the circuit.

Then if we adopt the convention that a Boolean function with $n$ inputs and $m$ outputs is in a quantum circuit class if its reversible version is, we clearly have $\mathbf{AC}^k \subseteq \mathbf{QAC}^k_{\text{wf}}$ and $\mathbf{ACC}^k \subseteq \mathbf{QACC}^k$. Thus we have

$$\mathbf{AC}^0 \subset \mathbf{ACC}^0[2] \subset \mathbf{ACC}^0 \subseteq \mathbf{QAC}^0_{\text{wf}} = \mathbf{QACC}^0[2] = \mathbf{QACC}^0$$

showing that $\mathbf{QAC}^0_{\text{wf}}$ and $\mathbf{QACC}^0[2]$ are more powerful than $\mathbf{AC}^0$ and $\mathbf{ACC}^0[2]$ respectively.

Interestingly, if $\mathbf{QAC}^0$ as we first defined it cannot do fanout, i.e. of $\mathbf{QAC}^0 \subset \mathbf{QAC}^0_{\text{wf}}$, then in a sense it fails to include $\mathbf{AC}^0$, since the fanout function from $\{0,1\}$ to $\{0,1\}^n$ is trivially in $\mathbf{AC}^0$. However, it is not clear whether it fails to include any $\mathbf{AC}^0$ functions with a one-bit output. On the other hand, if $\mathbf{QAC}^0$ can do fanout, it can also do parity and is greater than $\mathbf{AC}^0$, so either way $\mathbf{AC}^0$ and $\mathbf{QAC}^0$ are different. I am indebted to Pascal Tesson for pointing this out.

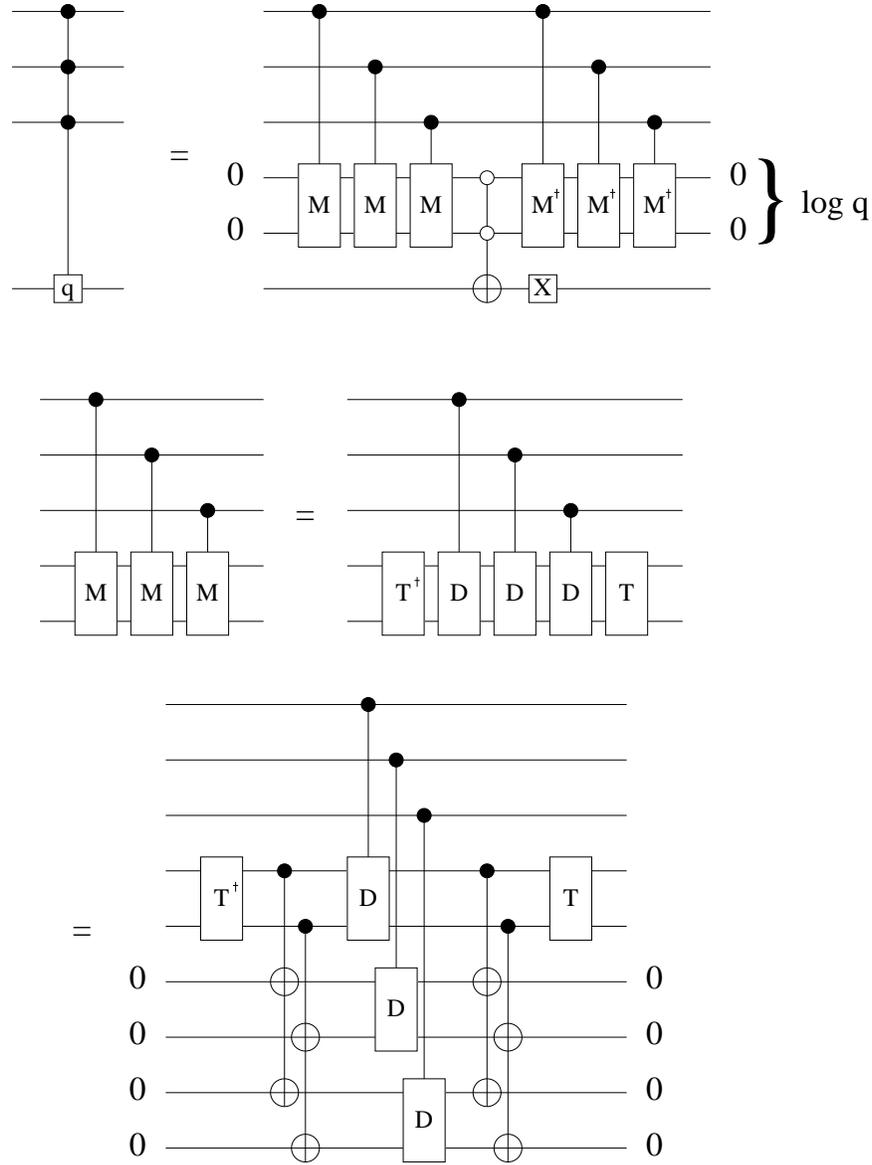

**Fig. 5.** Building a $\mathrm{MOD}_q$ gate. We choose a matrix $M$ on $k = \lceil \log q \rceil$ qubits such that $M^q = \mathbf{1}$, apply controlled-$M$ gates from the $n$ inputs to $k$ ancillae, perform an OR on the target qubit, and reverse the process. To parallelize this, we can diagonalize $M$ by writing it as $T^\dagger D T$, fan the $k$ qubits out into $n$ copies each using Proposition 1, and apply controlled-$D$ gates simultaneously from each input to a set of copies. The total depth depends on $q$ but not on $n$.

## 5 Conclusion

We have shown that fanout and parity are intimately related in quantum circuits, and that either of these makes it possible to construct a $\text{MOD}_q$ gate in constant depth for any $q$. Thus $\mathbf{QAC}^0_{\text{wf}} = \mathbf{QACC}^0[2] = \mathbf{QACC}^0$, and so $\mathbf{QAC}^0_{\text{wf}}$ and $\mathbf{QACC}^0[2]$ are more powerful than $\mathbf{AC}^0$ and $\mathbf{ACC}^0[2]$ respectively. This adds another case where quantum classes are provably more powerful than their classical cousins.

We suggest three questions for further work:

1. Is $\mathbf{QAC}^0 = \mathbf{QAC}^0_{\text{wf}}$? That is, can the fanout gate be constructed in constant depth when each qubit can only act as an input to one gate in each layer?
2. Is $\mathbf{QACC}^0[q] = \mathbf{QACC}^0$ for all $q$? That is, can $\text{MOD}_q$ gates, where $q$ is odd, generate fanout, parity, and $\text{MOD}_p$ gates for all other $p$?
3. Is $\mathbf{QAC}^0_{\text{wf}} = \mathbf{QTC}^0$? That is, can the techniques used here be extended to construct threshold gates in constant depth?

We conjecture that the answer to all these questions is 'no,' but quantum circuits can be surprising...

**Acknowledgments.** I thank Denis Thérien for organizing the 1999 McGill Workshop on Computational Complexity, and Pascal Tesson and Charles Bennett for helpful conversations. I also thank Molly Rose and Spootie the Cat for their support. This work was supported in part by NSF grant ASC-9503162.